\documentstyle[aps,prbbib,twocolumn,epsf]{revtex}

\begin{document}
\topmargin -1.4cm

\draft
\title{Excess Kondo resonance in a quantum dot device with normal 
and superconducting leads: the physics of Andreev-normal co-tunneling}

\author{Qing-feng Sun$^1$, Hong Guo$^1$, and Tsung-han Lin$^2$}

\address{$^1$Center for the Physics of Materials and Department
of Physics, McGill University, Montreal, PQ, Canada H3A 2T8.\\
$^2$State Key Laboratory for Mesoscopic Physics and
Department of Physics, Peking University, Beijing 100871, China}
%\date{}

\maketitle

\begin{abstract}
We report on a novel Kondo phenomenon of interacting quantum dots coupled 
asymmetrically to a normal and a superconducting lead. The effects of intradot 
Coulomb interaction and Andreev tunneling give rise to Andreev bound 
resonances. As a result, a new type of co-tunneling process which we term 
Andreev-normal co-tunneling, is predicted. At low temperatures, coherent 
superposition of these co-tunneling processes induces a Kondo effect in 
which Cooper pairs directly participate formation of a spin singlet, leading 
to four Kondo resonance peaks in the local density of states, and enhancing 
the tunneling current.
\end{abstract}
\pacs{72.15Qm, 73.40Gk, 72.15Nj}
%\newpage

The Kondo effect is a prototypical many-body correlation effect involving
interactions between a localized spin and free electrons.\cite{new1} Its recent 
observation in semiconductor quantum dots (QD)\cite{ref1,ref2,ref4} has 
generated a great deal of theoretical and experimental interests and provided 
rich understanding to many-body phenomena at the mesoscopic scale.\cite{addprl}
For a QD coupled to two normal (N) leads, the physical origin of the Kondo 
effect is now understood\cite{new1,ref1,ref4}. Consider a single spin degenerate 
level $\epsilon_d$ of the QD such that $\epsilon_d<\mu_N<\epsilon_d +U$, 
where $\mu_N$ is the chemical potential of the leads and $U$ the on-site e-e
interaction energy. An electron of either spin up or spin down which occupies 
$\epsilon_d$ cannot tunnel out of the QD because $\epsilon_d<\mu_N$; as well, 
an electron outside the QD cannot tunnel into it unless the on-site Coulomb 
energy $U$ is overcome. Therefore, the first-order tunneling process is 
Coulomb blockaded. However, due to Heisenberg uncertainty, the virtual 
higher-order co-tunneling events can still take place\cite{new1,ref1,ref4} by 
which the electron inside QD tunnels out followed by an electron with opposite 
spin tunneling into the QD, on a time scale $\sim \hbar/|\mu_N-\epsilon_d|$. 
As a consequence, 
the local spin is flipped. At low temperatures, the coherent superposition of 
all possible co-tunneling events gives rise to the Kondo effect in which the 
time-averaged spin in the QD is zero due to frequent spin flips: the whole 
system, QD plus leads, forms a spin singlet, and a very narrow Kondo peak 
located at $\mu_N$ arises in the local density of states (LDOS).

When one of the leads is a superconductor (S), another transport 
process---Andreev tunneling, will occur in the normal-superconductor 
interface in which an incident electron from the normal side is reflected as 
a hole while a Cooper pair is created in the superconductor. Andreev 
tunneling is very important\cite{ref5} because it determines transport 
properties of many mesoscopic superconducting-normal hybrid devices. It is 
therefore not surprising that the Kondo effect in N-QD-S hybrid systems
has attracted considerable attention\cite{ref7,ref8,ref9,ref10,ref11}.
So far, the focus on Kondo effects in N-QD-S devices has been on enhancement 
or reduction of conductance as compared to that of N-QD-N 
systems\cite{ref7,ref8,ref10}; and the emergence of sub-Kondo peaks in LDOS at 
$\pm\Delta$\cite{ref7,ref8} where $\Delta$ is the superconductor gap energy. 
In these studies Andreev tunneling precesses, in essence, happen alone
while the Cooper pairs do not participate the formation of spin-singlet. 
However, since the Kondo effect in a QD results from co-tunneling processes,
for a N-QD-S hybrid system, it is very natural to ask: are there 
co-tunneling processes consisting of one virtual Andreev tunneling and one 
virtual normal electron tunneling? If there are, can coherent superpositions 
of these Andreev-normal co-tunneling give rise to a Kondo effect? What are 
the consequences and characteristics of the Kondo effect induced this way?

It is the purpose of this letter to report our theoretical investigation on 
these issues. In contrast to previous work\cite{ref7,ref8,ref9,ref10,ref11},
we emphasize the possibility of virtual Andreev tunneling directly 
participating the co-tunneling process so that the Cooper pairs directly
participate formation of the spin-singlet: these physical processes
give rise to Kondo effect in the first place. Our results predicts a new 
co-tunneling process formed by an Andreev tunneling and a normal tunneling, 
and the superposition of this type processes induce four Kondo peaks in the 
LDOS.
 
We consider the standard model Hamiltonian\cite{ref13}
of a N-QD-S system $H = H_N + H_S + H_{QD} + H_T$ where $H_N = 
\sum_{k\sigma}\epsilon_{Nk}a^{\dagger}_{k\sigma}a_{k\sigma}$
and $H_S = \sum_{k\sigma}\epsilon_{Sk}b^{\dagger}_{k\sigma}b_{k\sigma} + 
\sum_k \left( \Delta b_{k\downarrow}b_{-k\uparrow} + H.c. \right)$ 
describe the normal lead and the superconducting lead, respectively,
in here we have set $\mu_S=0$.
$H_{QD}=\sum_{\sigma}\epsilon_d d_{\sigma}^{\dagger} d_{\sigma} +
Ud^{\dagger}_{\uparrow}d_{\uparrow}d^{\dagger}_{\downarrow}d_{\downarrow}$ 
models the QD with a single level having spin index 
$\sigma=\uparrow,\downarrow$ and intradot e-e Coulomb 
interaction $U$; $H_T=\sum_{k\sigma}[V_N a^{\dagger}_{k\sigma}d_{\sigma}
+V_S b^{\dagger}_{k\sigma}d_{\sigma} +H.c]$ denotes the tunneling part 
of the Hamiltonian.
The current from the normal lead flowing into the QD is calculated by
the standard Keldysh nonequilibrium Green's function theory, 
as ($\hbar=1$)\cite{ref13}:
\begin{equation}
I = -4e Im \int \frac{d\epsilon}{2\pi} \Gamma_N 
\left\{ f_N(\epsilon) {\bf G}^r(\epsilon)+\frac{1}{2} {\bf G}^<(\epsilon)
\right\}_{11}
\label{I1}
\end{equation}
where $\Gamma_{N/S}(\epsilon) \equiv 2\pi \sum_k |V_{N/S}|^2 
\delta(\epsilon-\epsilon_{N/S,k})$;
$f_N(\epsilon)$ is the Fermi distribution of the normal lead. 
The subscript ``11" means taking the ``11" element of the $2\times 2$ 
matrix. The entire analysis therefore falls on to derivations of retarded and 
Keldysh Green's functions ${\bf G}^r$ and ${\bf G}^<$ for the QD in the well 
known Nambu representation\cite{ref13}.

We have solved ${\bf G}^r(\epsilon)$ using the equation of motion method. 
Although this method is {\it quantitatively} less accurate in predicting
intensity of Kondo effect, it has been proven to provide correct 
{\it qualitative} physics at low temperatures\cite{ref15}, therefore it is 
sufficient for the purpose of this work. 
Then we obtain the matrix form of 
${\bf G}^r(\epsilon)$ as,\cite{add2}
$$
  \left( \begin{array}{ll} \epsilon-\epsilon_d - \Sigma^{(0)}_{11}  +
U A_1 B (\Sigma^{ab}_{11} +\Sigma^{(0)}_{S;12}A_2 \Sigma^{(b)}_{S;21}) 
         &
          -\Sigma^{(0)}_{S;12} - U A_1 B (\Sigma^{(b)}_{S;12}
                  + \Sigma^{(0)}_{S;12} A_2 \Sigma^{ab}_{22})  \\
          -\Sigma^{(0)}_{S;21} + U A_2 B (\Sigma^{(b)}_{S;21}
                  + \Sigma^{(0)}_{S;21} A_1 \Sigma^{ab}_{11}) &
          \epsilon+\epsilon_d - \Sigma^{(0)}_{22} - U A_2 B
(\Sigma^{ab}_{22} +\Sigma^{(0)}_{S;21}A_1 \Sigma^{(b)}_{S;12}) 
         \end{array}
   \right) 
    {\bf G}^r
      $$
\begin{equation}
   = \left( \begin{array}{ll}
       1+U A_1 B ( n_{\downarrow} + \Sigma^{(0)}_{S;12} A_2
                <d^{\dagger}_{\uparrow}d^{\dagger}_{\downarrow}> ), &
       U A_1 B ( <d_{\downarrow} d_{\uparrow}>
            -  \Sigma^{(0)}_{S;12} A_2 n_{\uparrow}  )  \\
       U A_2 B ( <d_{\uparrow}^{\dagger} d_{\downarrow}^{\dagger}>
            +  \Sigma^{(0)}_{S;21} A_1 n_{\downarrow}  ),     &
       1- U A_2 B ( n_{\uparrow} - \Sigma^{(0)}_{S;21} A_1
                <d_{\downarrow}d_{\uparrow}> )  
        \end{array} \right)
\label{Gr}
\end{equation}
where $A_i(\epsilon)$($i=1,2$) and $B(\epsilon)$ are defined as
$A_i(\epsilon) \equiv [\epsilon \mp \epsilon_d \mp U - 2\Sigma^{(0)}_{ii}
-\Sigma^{(1)}_{ii} ]^{-1}$;
$B(\epsilon)\equiv [1-\Sigma^{(0)}_{S;12} A_2 \Sigma^{(0)}_{S;21} A_1]^{-1}$;
and the self-energies ${\bf \Sigma}={\bf \Sigma}_N+{\bf \Sigma}_S$ are:
\begin{eqnarray}
& & {\bf \Sigma}^{(0)}_{S} (\epsilon) =\sum_k \frac{V^2_S}{E}
                   \left( \begin{array}{ll}
                           \epsilon_{Sk}^+  & -\Delta \\
                           -\Delta   &  \epsilon_{Sk}^-
                  \end{array} \right) , \nonumber \\
& & 
\Sigma^{(1)}_{S;ii} = \sum_k
\frac{V_S^2(\epsilon_{Sk}^{\mp}\mp 2\epsilon_d\mp U)}
{(\epsilon\mp 2\epsilon_d \mp U +i0^+)^2 -E_{Sk}^2 }\ \ \ ,\nonumber  \\
& & 
\Sigma^{(a)}_{S;ii}=\sum_k\frac{V_S^2[(\epsilon_{Sk}^{\mp}\mp 2\epsilon_d\mp U)
n_{k\downarrow} \mp \Delta p_i]}
{(\epsilon\mp 2\epsilon_d \mp U +i0^+)^2 -E_{Sk}^2 }\ , \nonumber \\
& &
{\bf \Sigma}^{(b)}_{S} = \sum_k \frac{V_S^2}{E} \left( \begin{array}{ll}
         \epsilon_{Sk}^+ n_{k\downarrow} +\Delta p_2  &
         2( \epsilon_{Sk}^+ p_1 - \Delta n_{k\downarrow} ) \nonumber \\
         2( - \epsilon_{Sk}^- p_2 - \Delta n_{k\uparrow} ) &
         \epsilon_{Sk}^- n_{k\uparrow} -\Delta p_1
\end{array} \right)\   \nonumber 
\end{eqnarray}
and $
\Sigma^{ab}_{ii} = \Sigma^{(a)}_{ii} + \Sigma^{(b)}_{ii} $, 
where index $i=1,2$. In these expressions 
$E\equiv \epsilon_{Sk}^+\epsilon_{Sk}^- -\Delta^2$,
$\epsilon_{Sk}^{\pm}\equiv \epsilon\pm\epsilon_{Sk}+i0^+$,
$E_{Sk}^2\equiv \Delta^2+\epsilon_{Sk}^2$, 
$n_{k\sigma} \equiv <b_{k\sigma}^{\dagger} b_{k\sigma}>$,
$p_1\equiv <b_{-k\downarrow}b_{k\uparrow}>$, and $p_2= p_1^*$. 
The self-energy $\Sigma_N$ for the coupling to the normal lead can easy be 
obtained from $\Sigma_S$ by setting $\Delta=0$ and substituting 
($V_S,\epsilon_{Sk},b_{k\sigma}$) by ($V_N,\epsilon_{Nk},a_{k\sigma}$). 
The quantity $n_{\sigma}$ in 
Eq.(\ref{Gr}) is the intradot electron occupation number 
of state $\sigma$; $<d_{\downarrow}d_{\uparrow}>$ and 
$<d_{\uparrow}^{\dagger}d_{\downarrow}^{\dagger}>$ are the pair 
correlation in the QD due to the well known proximity effect.
These quantities must be calculated self-consistently\cite{ref7,add1}. 
We emphasis that Eq.(\ref{Gr}) is suitable for arbitrary Coulomb interaction 
strength $U$ and superconducting gap $\Delta$. In the zero-gap limit so that
the superconducting lead becomes a normal lead, Eq.(\ref{Gr}) reduces to 
that of the N-QD-N system\cite{ref16}. 

Next, we solve the Keldysh Green's function ${\bf G}^<(\epsilon)$ which, for 
interacting systems, can not be obtained from the equation of motion without 
introducing additional assumptions. We follow the most commonly used ansatz 
for interacting lesser (greater) self-energy ${\bf \Sigma}^{\alpha}$
($\alpha=<$ and $>$), due to Ng\cite{ref19}, but we generalize this ansatz 
to mesoscopic hybrid systems in the following way,
\begin{equation}
{\bf \Sigma}^{\alpha} = \frac{1}{2}\left[ 
{\bf \Sigma}^{\alpha}_0 ({\bf \Sigma}^r_0-{\bf \Sigma}^a_0)^{-1}{\bf X} 
+ {\bf X}({\bf \Sigma}^r_0-{\bf \Sigma}^a_0)^{-1} {\bf \Sigma}^{\alpha}_0 
\right] 
\label{Sigmaless}
\end{equation}
%\vspace{40mm}
where ${\bf \Sigma}_0$ is the exact self-energy for noninteracting system. 
Then, from ${\bf \Sigma}^<-{\bf \Sigma}^>= {\bf \Sigma}^r - {\bf \Sigma}^a$, 
it is easy to determine ${\bf X}={\bf \Sigma}^r - {\bf \Sigma}^a$. 
This ansatz has several advantages: (i) it is exact both in equilibrium 
(the bias $V_b=0$) and in noninteracting limit ($U=0$); (ii) when $\Delta=0$, it is 
consistent with the original ansatz of Ng\cite{ref19}; (iii) the current 
conservation is automatically satisfied; (iv) it guarantees the exact 
relation of matrix elements $G_{12}^<(\epsilon)=-G_{21}^{<*}(\epsilon)$, 
\vspace{40mm}
which in turn guarantees $<d_{\downarrow}d_{\uparrow}>=<d_{\uparrow}^{\dagger}
d_{\downarrow}^{\dagger}>^*$; (v) $G_{11}^<(\epsilon)$ and 
$G_{22}^<(\epsilon)$ automatically becomes purely imaginary. In this regard, 
we note that Ref.\onlinecite{ref11} proposed to use only the first term of 
Eq.(\ref{Sigmaless}) as ${\bf \Sigma}^<$. At least for our purpose this 
choice is not correct because our matrices in (\ref{Sigmaless}) are 
non-diagonal therefore cannot be permuted, and more importantly, it generates 
result violating points (iv) and (v). Finally, using
(\ref{Sigmaless}), ${\bf G}^<$ is obtained from the Keldysh equation 
${\bf G}^<= {\bf G}^r {\bf \Sigma}^< {\bf G}^a$ where
${\bf G}^a =({\bf G}^r)^{\dagger}$. With Green's functions ${\bf G}^r$ 
and ${\bf G}^<$, from Eq.(\ref{I1}) the current $I$ is calculated immediately. 

To gain physical insights to the analytical result, in the rest of the paper
we discuss them numerically. In our numerical calculations, we assume 
square bands of width $2W$ so that $\Gamma_{N/S}(\epsilon)= \Gamma_{N/S}
\theta(W-|\epsilon|)$, with
$W=1000\gg max(k_BT,V_b,\Gamma, \Delta)$. 
We emphatically investigate the case when $\Delta>\Gamma_S$ and with
asymmetrical barriers, $\Gamma_S>\Gamma_N$. In this case, an electron with 
energy $|\epsilon|<\Delta$ in the QD undergoes multiple Andreev reflections 
before it decays to the normal lead, and it cannot decay into the 
superconductor due to the gap. These multiple reflections give rise to 
Andreev bound states in the QD which are indicated by the peaks in the 
LDOS\cite{add4}. Fig.(1a) shows LDOS at a high temperature $T=0.5$. If $U=0$, 
two Andreev bound states emerge at $\pm\sqrt{\epsilon^2_d+
\Gamma^2_S/4}$ (when $\Delta\gg\Gamma_S\gg\Gamma_N$). When $U$ becomes finite, 
due to a competition of intradot Coulomb interaction and Andreev tunneling, 
each $U=0$ Andreev bound state is split into two sub-states an energy 
$U$ apart. As a result, four Andreev bound state peaks emerge in the LDOS 
each with half-width set by $\Gamma_N$. Note that for large interactions 
$U\rightarrow\infty$, Andreev reflections are Coulomb blockaded therefore 
only two ordinary resonance peaks at $\epsilon_d$ and $\epsilon_d+U$ can
be detected in LDOS.

Next, we drop the temperature to $T=0.005$ so that Kondo effect can be 
investigated, the data shown in Fig.(1b). In LDOS, the four broad peaks 
correspond to the four Andreev bound states just discussed. On top of them,
four additional narrow peaks emerge. These narrow peaks only exist at low 
temperature and they have the typical character of Kondo effect:
their peak heights greatly depend on $T$ and they heighten with reducing 
$T$. The positions of these Kondo peaks are at $\epsilon=\mu_N$, $-\mu_N$,
$2\epsilon_d+U-\mu_N$, and $-2\epsilon_d-U+\mu_N$.
Note that two of them are dependent on $\epsilon_d$ which is qualitatively
different from the conventional Kondo effect.
% independent of $T$. 
Moreover, the Kondo peaks at $-\mu_N$ and $-2\epsilon_d-U+\mu_N$ are outside
the Kondo region of a normal system. Thus, instead of one Kondo peak, we now 
have three ``excess Kondo peaks''. At bias $V_b=0.7$, the two Kondo peaks at 
$2\epsilon_d+U-\mu_N$ and $-2\epsilon_d-U+\mu_N$ overlap each other, 
therefore only three peaks are observed (see Fig.(1b)). As an extremely 
strict confirmation of our analytical derivations and numerical calculations, 
for all cases we have checked that $\int d\epsilon [LDOS(\epsilon)]=2$
which indicates that there are two states in the QD.
 
Where do the excess Kondo peaks originate? They certainly cannot be due to
first-order tunneling processes because 
Andreev bound states do not align with $\mu_N$.
% $\epsilon_d<\mu_N, \Delta$ and 
% $\epsilon_d +U >\mu_N, -\Delta$. 
In addition, although direct Andreev 
tunneling does occur, their effect is to evolve the intradot level 
$\epsilon_d$ into the Andreev bound states, {\it i.e.} to give the four broad 
peaks in the LDOS as discussed in the above.
They cannot give rise 
to the Kondo effect simply because this process can not flip local spin. 

Our investigation suggests that the excess Kondo peaks originate from an 
interesting co-tunneling process not discovered before, which is indicated 
by Fig.(2). First, the conventional Kondo peak at $\epsilon=\mu_N$ originates 
from the normal co-tunneling event between the QD and the normal lead depicted
in Fig.(2a). Note that this process does not induce any net current. 
Second, the excess Kondo peak at $\epsilon=-\mu_N$ is from a Andreev-normal 
co-tunneling process described in Fig.(2b-d). To start, an electron 
with spin-up, for example, occupies the QD (Fig.2b). Then a down-spin electron 
in the normal lead with energy $\sim \mu_N$ can tunnel into QD and reach 
the QD-S interface causing an Andreev reflection by which a hole is reflected 
back, Fig.(2c). If $\Gamma_N \geq \Gamma_S$, this hole can easily tunnel into 
the normal lead and neutralizes an up-spin electron. This is a real Andreev 
process in which two electrons with opposite spin in the normal lead are 
annihilated concomitant to the creation of a Cooper pair in the superconductor,
leaving the occupancy and the spin of the QD unchanged. 
However, when $\Gamma_S>\Gamma_N$ which is our concern, it is more difficult for this 
hole to tunnel into the normal lead, the probability is increased for this 
hole to combine with the original up-spin electron of the QD (Fig.2c). Note 
that this is a {\it virtual} Andreev tunneling process: after it the system 
is in a high-energy virtual state which can only exist for a timescale 
$\sim \hbar/|\mu_N+\epsilon_d|$. Closely following this virtual Andreev 
process, another virtual process involving a normal tunneling event can 
occur where a down-spin electron with energy $\sim -\mu_N$ tunnels into the QD.
As consequences of this virtual Andreev and virtual normal co-tunneling 
process, two electrons with same spin in the normal lead are annihilated, a 
Cooper pair is created in the superconductor, the spin of the QD is flipped 
and a net current flows through the QD. At low temperatures, a coherent 
superposition of all possible co-tunneling processes of this kind produces 
the Kondo effect. The entire system, QD plus leads and including the Cooper 
pair, forms a spin singlet. The local spin is effectively screened and two 
very narrow Kondo peaks emerge at $\epsilon=\mu_N$ and $-\mu_N$ in the LDOS. 
The new physics here is that Andreev tunneling can directly participate 
co-tunneling processes, in other words the Cooper pair can directly 
participate formation of the spin singlet. 

Third, the excess Kondo peak at $2\epsilon_d+U-\mu_N$ is from the Andreev 
assisted co-tunneling process. An Andreev event involves a two-particle 
tunneling process, hence the electron number in QD can fluctuate by two. 
In particular, when $\Gamma_S>\Gamma_N$, Andreev tunneling occurs more 
frequently than normal tunneling, therefore the probability for QD to have 
zero- or two-occupancy is greatly enhanced. This induces the following 
co-tunneling process. When QD is empty, an electron with energy $\sim \mu_N$ 
in the normal lead can tunnel into the QD state $\epsilon_d$, following 
closely by another electron having opposite spin and energy 
$\sim 2\epsilon_d+U-\mu_N$ tunneling on to the state $\epsilon_d+U$. 
If the QD begins with a two-occupancy, the inverse process can occur. These 
co-tunneling processes induce the Kondo peak at $2\epsilon_d +U-\mu_N$.
Notice that the electron fluctuation is two in these co-tunneling process, 
so it only occurs under the assistance of Andreev tunneling. Finally, the 
Kondo peak at $-2\epsilon_d-U+\mu_N$ originates from a combination of the 
processes discussed in this and the last paragraph. 

The above picture clearly indicates the physical process behind the excess 
Kondo peaks. Is the tunneling current enhanced by these excess Kondo 
resonances? The inset of Fig.(1c) plots current versus $\epsilon_d$ at bias 
$V_b=0.5$. A finite bias makes the conventional Kondo peak at $\mu_N$ not 
to align with $\mu_S$, so that this peak alone contributes weakly to current. 
At low temperatures (solid curve), our results suggest that the current is 
indeed enhanced in the region $-U$ to $0$ by the excess Kondo effect because 
the Andreev-normal co-tunneling does generate net current flow. Under what 
conditions can we observe the excess Kondo resonances? (i). When 
$\Gamma_N\geq \Gamma_S$, no excess Kondo peaks can be obtained and 
only the conventional Kondo peak at $\mu_N$ emerges (inset of Fig.1b).
This is because the 
normal tunneling at the N-QD interface is frequent which diminishes the 
virtual Andreev tunneling and the two electron fluctuation processes 
necessary for the excess Kondo peaks, as well suppresses the formation of 
Andreev bound states. (ii). When $U\rightarrow\infty$, all excess Kondo peaks 
disappear, because at large $U$ the timescale (lifetime) of the virtual 
two-occupancy state, $\sim \hbar/|U+\epsilon_d-\mu_N|$,
goes to zero thereby greatly suppressing the Andreev tunneling (including
the virtual process).
This is the qualitative difference between the excess Kondo peaks discovered 
in this work and the Kondo peaks at $\pm\Delta$ when $\Delta \sim \Gamma$: 
the latter survive the $U\rightarrow\infty$ limit because their physical origin
is from co-tunneling between the QD and the superconducting quasi-particle 
spectrum and does not involve direct participation of Andreev tunneling 
process. (iii). When $\mu_N<\epsilon_d$, all Kondo peaks disappear and four 
broad peaks due to Andreev bound states still exist as shown in Fig.(1c). 
Although $-\mu_N$ is still between $\epsilon_d$ and $\epsilon_d+U$, the excess 
Kondo peak at $-\mu_N$ nevertheless disappears because when $\mu_N<\epsilon_d$,
the Fermi surface is below $\epsilon_d$ which forbids all co-tunneling process
from happening.

In summary, for a N-QD-S device satisfying the condition 
$\Delta>\Gamma_S>\Gamma_N$, excess Kondo peaks are predicted in the local 
density of states which enhances the tunneling current. These excess Kondo 
peaks are due to an interesting co-tunneling process involving an Andreev 
tunneling from the QD-S interface and a normal tunneling from the N-QD 
interface. This new co-tunneling phenomenon clearly demonstrates the rich 
many-body physics of the hybrid system in which the formation of a spin
singlet not only can be due to the many free electrons, but also can be due
to a combined effect of free electrons and the many Cooper pairs in the 
superconducting lead.

{\bf Acknowledgments:}
We gratefully acknowledge financial support from NSERC of Canada, FCAR of
Quebec (Q.S., H.G), and NSF of China (T.L.).

%\bigskip

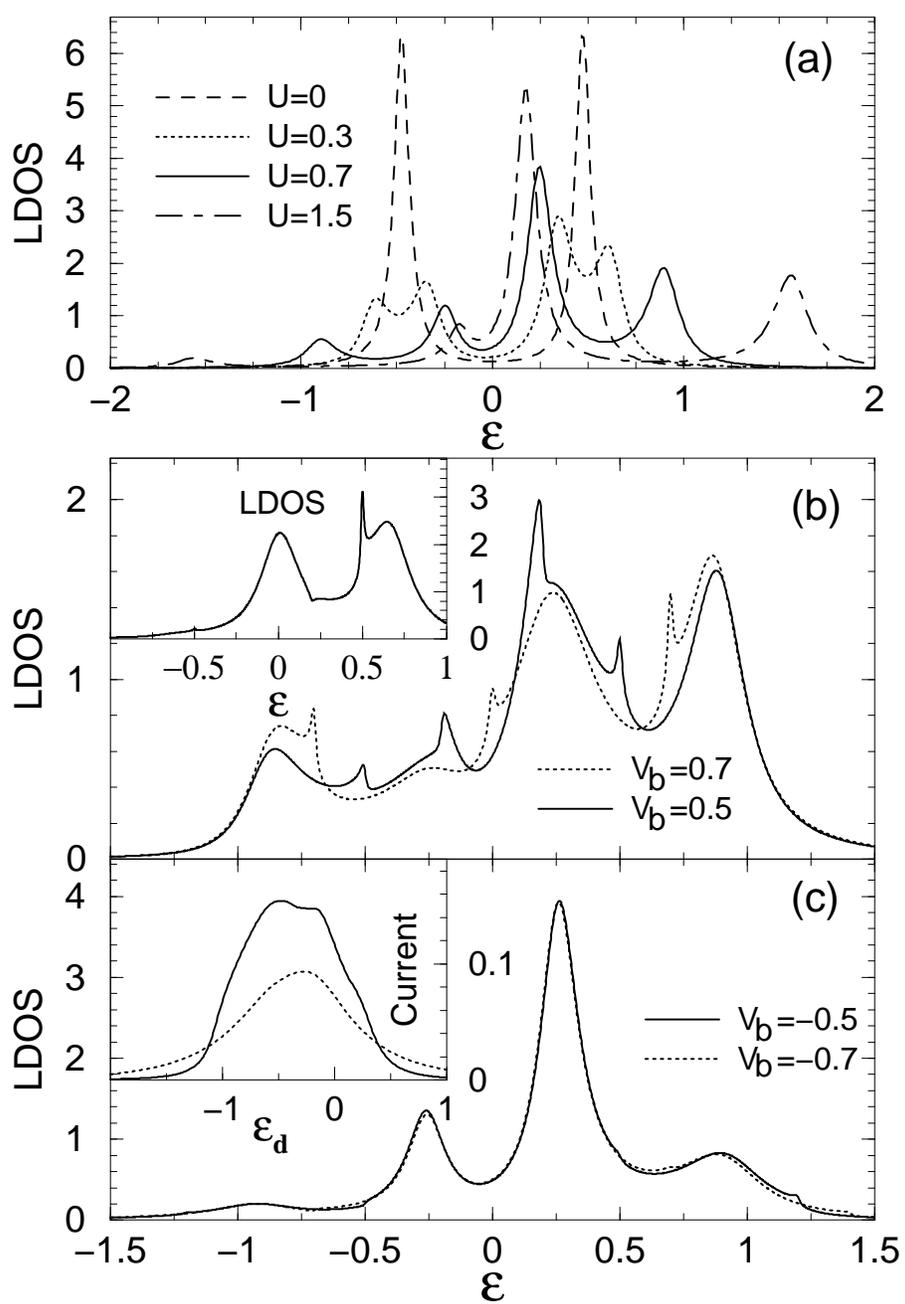
\begin{figure}
\caption{
(a) LDOS vs $\epsilon$ at different $U$ for a high temperature $T=0.5$, 
with parameters $V_b=0$, $\Gamma_N=0.1$. (b) and (c) are LDOS at different 
bias $V_b$ for a low temperature $T=0.005$, and parameters $U=0.7$, 
$\Gamma_N=0.15$. Other parameters are $\epsilon_d=0$, $\Gamma_S=1$, 
$\Delta=10$. The inset of (b) shows LDOS vs $\epsilon$ for symmetric 
barriers with $\Gamma_N=\Gamma_S=0.15$. The inset in (c) shows current vs 
$\epsilon_d$ at $T=0.5$ (dotted curve) and $T=0.005$ (solid curve).
Other parameters of insets are same as those of the solid curve in (b).
}
\label{fig1}
\end{figure}

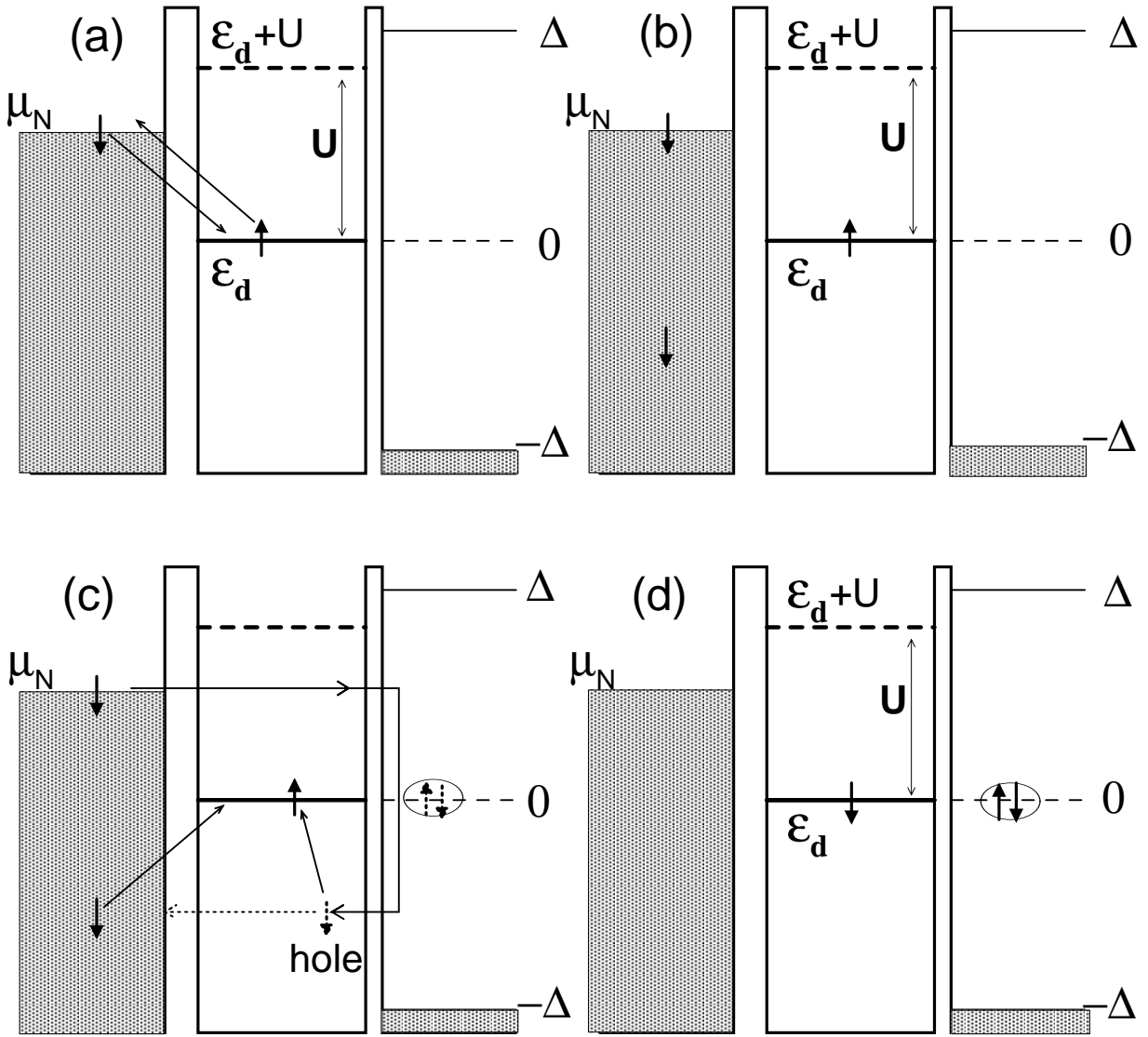
\begin{figure}
\caption{
Schematic plots showing the co-tunneling process. (a) Conventional normal
co-tunneling in the N-QD interface. (b), (c), and (d) are the Andreev-normal 
co-tunneling. (b) the initial state; (c) the virtual tunneling event of 
the Andreev-normal co-tunneling; (d) the final state. The dotted arrow
in the QD of (c) represents a hole, other arrows represent electron. 
}
\label{fig2}
\end{figure}


\begin{references} 
\bibitem{new1}
L. P. Kouwenhoven and L. Glazman, Physics World {\bf Jan.}, 33 (2001).

\bibitem{ref1}  
S. M. Cronenwett, T. H. Oosterkamp and L. P. Kouwenhoven, Science {\bf 281}, 540 (1998). 

\bibitem{ref2}  
T. Inoshita, Science {\bf 281}, 526 (1998); 
W. G. van der Wiel {\sl et al}, Science {\bf 289}, 2105 (2000);
D. Goldhaber-Gordon {\sl et al}, Nature {\bf 391}, 156 (1998);
Phys. Rev. Lett. {\bf 81}, 5225 (1998).

\bibitem{ref4}  
S. Sasaki {\sl et al}, Nature {\bf 405}, 764 (2000).

\bibitem{addprl}
I. Affleck and P. Simon, Phys. Rev. Lett. {\bf 86}, 2854 (2001);
M. Eto and Y. V. Nazaro, {\sl ibid} {\bf 85}, 1306 (2000);
B. R. Bulka and P. Stefanski, {\sl ibid} {\bf 86}, 5128 (2001).

\bibitem{ref5}  
B. J. van Wees and H. Takayanagi, in {\sl Mesoscopic
Electron Transport}, edited by L. L. Sohn {\sl et al},
(Kluwer, Dordrecht, 1997).

\bibitem{ref7}
J. C. Cuevas, A. L. Yeyati and A. Martin-Rodero, Phys. Rev. B {\bf 63}, 094515 (2001).

\bibitem{ref8}  
A. A. Clerk {\sl et al}, Phys. Rev. B {\bf 61}, 3555 (2000).

\bibitem{ref9}  
P. Schwab and R. Raimondi, Phys. Rev. B {\bf 59}, 1637 (1999).

\bibitem{ref10}  
K. Kang,  Phys. Rev. B {\bf 58}, 9641 (1998).
 
\bibitem{ref11} 
R. Fazio and R. Raimondi, Phys. Rev. Lett. {\bf 80}, 2913 (1998);
{\sl ibid} {\bf 82}, 4950(E) (1999).
  
\bibitem{ref13}  
Q. Sun, J. Wang and T. Lin, Phys. Rev. B {\bf 62}, 648 (2000).

\bibitem{ref15}  
Y. Meir, N. S. Wingreen and P. A. Lee, Phys. Rev. Lett. {\bf 70}, 2601 (1993).

\bibitem{add2}   
The algebraic details and other results will be presented elsewhere:
Q. Sun {\sl et al}. (unpublished). In our analysis, the decoupling approximation is: 
$ <\{Y X_1 X_2, d^{\dagger}_{\sigma'}\}> =
<X_1 X_2> <\{ Y,d^{\dagger}_{\sigma'}\}> $,  where $X_{1/2}$ represent the 
operator $a_{k\sigma}$, $a_{k\sigma}^{\dagger}$,  $b_{k\sigma}$, and
$b_{k\sigma}^{\dagger}$; $Y$ is $d_{\sigma}$ and $d_{\sigma}^{\dagger}$. 
 
\bibitem{add1} 
The self-consistent equations are:  $n_{\uparrow}=Im\int\frac{d\epsilon}{2\pi}
G^<_{11}(\epsilon)$, $1-n_{\downarrow}= 
Im\int\frac{d\epsilon}{2\pi} G^<_{22}(\epsilon)$,
$<d_{\downarrow}d_{\uparrow}>=-i \int\frac{d\epsilon}{2\pi} 
G^<_{12}(\epsilon)$, and $<d_{\uparrow}^{\dagger}d_{\downarrow}^{\dagger}>
=-i\int\frac{d\epsilon}{2\pi} G^<_{21}(\epsilon)$. 
 
\bibitem{ref16}  
Y. Meir, N. S. Wingreen and P. A. Lee, Phys. Rev. Lett. {\bf 66}, 3048 (1991).

\bibitem{ref19} 
T.-K. Ng, Phys. Rev. Lett. {\bf 76}, 487 (1996). 

\bibitem{add4} 
$LDOS(\epsilon)=-Im [G_{11}^r(\epsilon) +G_{22}^r(-\epsilon)]/\pi$.
 
\end{references}
\end{document}